\documentclass[twocolumn,showpacs,preprintnumbers,amsmath,amssymb]{revtex4}

\usepackage{graphicx}
\usepackage{dcolumn}
\usepackage{bm}
\usepackage{amscd}
\usepackage{amssymb}
\usepackage{amsfonts}
\usepackage{amsmath}
\usepackage{mathrsfs}

\begin{document}


\title{Mechanical similarity as a generalization of scale symmetry}

\author{E. Gozzi}%
 \email{gozzi@ts.infn.it}
  \author{D. Mauro}
 \email{mauro@ts.infn.it}
\affiliation{%
Department of Theoretical Physics, 
	     University of Trieste, \\
	     Strada Costiera 11, Miramare-Grignano, 34014 Trieste, Italy,\\
	     and INFN, Trieste, Italy
}%


\date{\today}

\begin{abstract}
In this paper we study the symmetry known \cite{Landau} as mechanical similarity (LMS) and present for any monomial potential. We analyze it in the framework of the Koopman-von Neumann formulation of classical mechanics and prove that in this framework the LMS can be given a canonical implementation. We also show that the LMS is a generalization of the scale symmetry which is present only for the inverse square potential. Finally we study the main obstructions which one encounters in implementing the LMS at the quantum mechanical level. 
\end{abstract}

\maketitle

\section{Introduction}

We know that in classical statistical mechanics the probability densities in phase space $\rho(\vec{r},\vec{p},t)$ evolve with the Liouville equation: 
\begin{equation}
\displaystyle i\frac{\partial}{\partial t} \rho(\vec{r},\vec{p},t)=\hat{\cal{\tilde{H}}}\rho(\vec{r},\vec{p},t), \label{zero}
\end{equation}
where $\hat{\cal\tilde{H}}$ is the so called Liouville operator, which is built out of the Hamiltonian $H(\vec{r},\vec{p})$ as follows:
\begin{equation}
\hat{\cal \tilde{H}}=-i\vec{\partial}_{p} H(\vec{r},\vec{p})\cdot\vec{\partial}_{{r}}+i\vec{\partial}_{{r}}H(\vec{r},\vec{p})\cdot \vec{\partial}_{{p}}. \label{liouv}
\end{equation}
In \cite{koop} Koopman and von Neumann replaced the space of probability densities $\rho(\vec{r},\vec{p})$ with a Hilbert space of states $|\psi, t\rangle$. Furthermore, they postulated for $|\psi,t\rangle$ the following evolution: 
\begin{equation}
\displaystyle i\frac{\partial}{\partial t}|\psi,t \rangle =\hat{\cal H} |\psi,t \rangle,
\; \textrm{where} \; \hat{\cal H}=\vec{\lambda}_r\cdot \vec{\partial}_pH-\vec{\lambda}_p\cdot \vec{\partial}_rH. \label{eqmot}
\end{equation}
In  the previous equation $\vec{r}$, $\vec{p}$, $\vec{\lambda}_r$, $\vec{\lambda}_p$ are operators \cite{gozzi} whose only non-zero commutators are the following \footnote{The reader should not be bothered by the fact that in this formalism $[r_i,p_j]=0$ because, after all, we are doing classical mechanics.}:
\begin{equation}
\displaystyle \left[r_i, \lambda_{r_j}\right] =i\delta_{ij}, \qquad  \left[p_i, \lambda_{p_j}\right]=i\delta_{ij}.  \label{conjugate}
\end{equation}
From the previous equation we see that $\vec{\lambda}_r$ and $\vec{\lambda}_p$ are canonically conjugated to $\vec{r}$ and $\vec{p}$. In particular, if we choose the representation in which $\vec{r}$ and $\vec{p}$ are operators of multiplication, then the 
$\vec{\lambda}$ become the following operators of derivation: 
\begin{displaymath}
\displaystyle \vec{\lambda}_r=-i\vec{\partial}_r, \qquad \vec{\lambda}_p=-i\vec{\partial}_p.
\end{displaymath}
In this representation the abstract vectors $|\psi, t\rangle$ become functions of $\vec{r}$ and $\vec{p}$ and the abstract equation of motion (\ref{eqmot}) becomes exactly the Liouville equation of motion for the state $\psi(\vec{r},\vec{p})$:
\begin{equation}
\displaystyle i\frac{\partial}{\partial t}\psi(\vec{r},\vec{p},t)  =\hat{\cal\tilde{H}} \psi(\vec{r},\vec{p},t).  \label{eqpsi}
\end{equation}
In the previous formula $\hat{\cal \tilde{H}}$ is just the Liouville operator of Eq. (\ref{liouv}). The equation of evolution of the probability density (\ref{zero}) can be easily derived from the equation of motion (\ref{eqpsi}) and from the other main postulate of the KvN formulation, i.e. that the probability densities $\rho$ are the modulus square of the $\psi$: $\rho(\vec{r},\vec{p})=|\psi(\vec{r},\vec{p})|^2$.

It is clear from all this that we can choose other representations in the KvN Hilbert space. For example we can choose to represent the states $|\psi,t\rangle$ over the basis given by the eigenstates of $\vec{r}$ and $\vec{\lambda}_p$. In this case the KvN states  become $\psi(\vec{r},\vec{\lambda}_p)$. They evolve with the equation of motion (\ref{eqmot}) or via the following kernel of propagation: \cite{scale}: 
\begin{eqnarray*}
&& \displaystyle \langle \vec{r}, \vec{\lambda}_p,\tau| \vec{r}_{\scriptscriptstyle 0}, \vec{\lambda}_{p_0},0 \rangle =
\int {\mathscr D}^{\prime\prime}\vec{r}\,{\mathscr D}\vec{p}\,{\mathscr D}\vec{\lambda}_r{\mathscr D}^{\prime\prime}\vec{\lambda}_p\nonumber\\
&& \; \exp\left[i \int \textrm{d}t \left(\vec{\lambda}_r\cdot \dot{\vec{r}}- \vec{p}\cdot \dot{\vec{\lambda}}_p-
\vec{\lambda}_r\cdot \vec{\partial}_pH+\vec{\lambda}_p\cdot\vec{\partial}_r H \right) \right],
\end{eqnarray*}
where the double prime in ${\mathscr D}^{\prime\prime}$ indicates that the path integral is over paths with fixed end points.
In particular, if we consider a Hamiltonian of the form $\displaystyle H=p^2/2+V\left(\vec{r}\,\right)$ we get:
\begin{eqnarray*}
&&\displaystyle \langle \vec{r}, \vec{\lambda}_p,\tau| \vec{r}_{\scriptscriptstyle 0}, \vec{\lambda}_{p_0},0 \rangle 
=\int {\mathscr D}^{\prime\prime}\vec{r}\,{\mathscr D}\vec{p}\,
{\mathscr D}\vec{\lambda}_r{\mathscr D}^{\prime\prime}\vec{\lambda}_p\nonumber \\
&& \quad \displaystyle \exp\left[i \int \textrm{d}t \left(\vec{\lambda}_r\cdot \dot{\vec{r}}- \vec{p}\cdot \left(\dot{\vec{\lambda}}_p+
\vec{\lambda}_r\right)+\vec{\lambda}_p\cdot\vec{\partial}_r V \right) \right]. \end{eqnarray*}
Performing above the functional integral over $\vec{p}$ we get a functional Dirac delta $\delta(\dot{\vec{\lambda}}_p+\vec{\lambda}_r)$. This means that we can perform also the functional integral over $\vec{\lambda}_r$ by replacing everywhere $\vec{\lambda}_r$ with $-\dot{\vec{\lambda}}_p$. In this way we can integrate away the canonical momenta $\vec{r}$ and $\vec{\lambda}_p$ to get the following path integral over the configurational variables:
\begin{eqnarray}
&&\displaystyle \langle \vec{r}, \vec{\lambda}_p,\tau| \vec{r}_{\scriptscriptstyle 0}, \vec{\lambda}_{p_0},0 \rangle 
=\int {\mathscr D}^{\prime\prime}\vec{r}\,{\mathscr D}^{\prime\prime}\vec{\lambda}_p\, \nonumber \\
&& \quad \exp\left[i \int \textrm{d}t \left(-\dot{\vec{\lambda}}_p\cdot \dot{\vec{r}}+\vec{\lambda}_p\cdot\vec{\partial}_r V(\vec{r}\,) \right) \right]. \label{final}
\end{eqnarray}
This is the main tool we will use in the next sections to study a 
symmetry called \cite{Landau} mechanical similarity.
We will indicate it with the acronym LMS for {\it L}andau {\it M}echanical {\it S}imilarity even if most probably it was introduced much before Landau. We will call it that way also to distinguish it from another similar symmetry (see Sec. V of Ref. \cite{deotto}). The LMS, which in classical mechanics holds for every monomial potential, turns out to be a natural generalization of the standard scale symmetry analyzed in \cite{scale}: the only difference is that in the LMS the variables are not transformed according to their physical dimensions like in the scale transformations. We will also prove in Secs. II and III that, while the scale symmetry can be implemented as a canonical transformation both in the standard phase space formulation of classical mechanics and in the KvN extended space, the LMS can be implemented as a {\it canonical} symmetry {\it only} in the enlarged KvN space. This fact suggests that the LMS may be more easily implementable at the quantum level if we first manage to formulate also quantum mechanics in the KvN space. This had already been done in Ref. \cite{map}. Unfortunately, as we will show in Secs. IV and V, there are obstructions in implementing the LMS at the quantum level not only in the standard formulation of quantum mechanics but also within the KvN space. This suggests that the LMS is a symmetry peculiar of classical mechanics but which cannot be realized at the quantum level. For this reason we think that the LMS could play a role in the study of the interplay between classical and quantum mechanics. Finally, in Sec. VI we make a comparison between our approach and the one of Ref. \cite{cal} on Newton-equivalent Hamiltonians. 


\section{A generalization of the scale symmetry}

For a generic monomial potential $\displaystyle V(\vec{r}\,)=g\frac{r^n}{n}$ the weight of the path integral (\ref{final}) becomes: 
\begin{equation}
\displaystyle \widetilde{\cal S}\equiv \int \textrm{d}t \left( -\dot{\vec{\lambda}}_p \cdot \dot{\vec{r}}+g \, r^{n-2} \vec{\lambda}_p\cdot \vec{r} \, \right). \label{gentilde}
\end{equation}
Let us now suppose we perform an infinitesimal rescaling of the time variable $\delta t=-\tilde{\alpha} t$. From (\ref{gentilde}) we see that, differently than in the standard action $\displaystyle S=\int \textrm{d}t \left(\dot{r}^{2}/2-gr^n/n\right)$, we can act not only on $\vec{r}$ but also on $\vec{\lambda}_p$ to get an invariance of the weight of the classical path integral (\ref{final}). It is easy to prove that the following transformations:
\begin{equation}
\delta \vec{r}=-\frac{2\tilde{\alpha}}{2-n}\, \vec{r}, \quad  \delta \vec{\lambda}_p =\frac{n\tilde{\alpha}}{2-n}\, \vec{\lambda}_p, \quad \delta t=-\tilde{\alpha} t \label{transformations}
\end{equation}
leave unchanged the $\widetilde{\cal S}$ of Eq. (\ref{gentilde}), so they are a symmetry for classical mechanics in the KvN formalism. Of course, these transformations depend explicitly on the exponent $n$ of the monomial potential that we are taking into account. 
For $n=-2$ we have an inverse square potential and the transformations (\ref{transformations}) reproduce exactly the scale transformations analyzed in \cite{scale}. In this sense we can say that Eq. (\ref{transformations}) is a generalization of the scale symmetry. It is well known that in the scale symmetry $\vec{r}$ transforms according to its ``physical" dimensions \cite{scale}. This is not the case anymore for the transformations in (\ref{transformations}) but nevertheless, the transformations (\ref{transformations}) are an invariance for classical mechanics.
If we apply Noether's theorem and use the definitions of the momenta canonically conjugated to $\vec{r}$ and $\vec{\lambda}_p$, i.e. $\vec{\lambda}_r=-\dot{\vec{\lambda}}_p$ and $\vec{p}=\dot{\vec{r}}$, see Eq. (\ref{gentilde}), then we get the following charge which is conserved in the enlarged KvN space:
\begin{eqnarray}
\displaystyle {\cal D}&=&t \hat{\cal H}-\frac{1}{2-n}\left(\vec{\lambda}_r\cdot\vec{r}+\vec{r}\cdot\vec{\lambda}_r\right)\nonumber \\
&& -\frac{n}{2(2-n)}\left(\vec{\lambda}_p\cdot\vec{p}+\vec{p}\cdot\vec{\lambda}_p\,\right). \label{mscharge}
\end{eqnarray}
In the previous formula we have symmetrized $\vec{r}$ and $\vec{\lambda}_r$, $\vec{p}$ and $\vec{\lambda}_p$, to have a Hermitian charge under the standard scalar product in the KvN Hilbert space \cite{DGM}:
\begin{equation}
\displaystyle \langle \psi| \tau\rangle=\int \textrm{d}\vec{r} \,\textrm{d}\vec{p}\, \psi^*(\vec{r},\vec{p}\,)\,\tau(\vec{r},\vec{p}\,). \label{scprod}
\end{equation}

Before going on, let us analyze two particular cases: first of all, let us take a harmonic oscillator, i.e. $n=2$. In the limit $n\to 2$ the coefficients in front of the round brackets of Eq. (\ref{mscharge}) tend to become equal and much bigger than the first term $t {\cal H}$. So in the case of a harmonic oscillator the charge ${\cal D}$ becomes roughly:
\begin{equation}
\displaystyle  {\cal D} \propto
\vec{\lambda}_r\cdot\vec{r}+\vec{p}\cdot\vec{\lambda}_p.  \label{msosc}
\end{equation}
It is easy to prove that this charge commutes with the Liouvillian associated with a harmonic oscillator\break $\hat{\cal H}=\vec{\lambda}_r\cdot \vec{p}-\vec{\lambda}_p\cdot \vec{r}$ and, being independent of $t$, it is conserved. This same charge plays an important role in one of 't Hooft's papers on the derivation of quantum mechanics from dissipative deterministic systems \cite{hooft}. As a second particular case, let us consider the inverse square potential for which $n=-2$. In this case the conserved charge of Eq. (\ref{mscharge}) reduces to the dilation charge that we found in Ref. \cite{scale}:
\begin{displaymath}
\displaystyle {\cal D}=t \hat{\cal H} +\frac{1}{2}\left(\vec{\lambda}_p\cdot\vec{p}-\vec{\lambda}_r\cdot\vec{r}\right),
\end{displaymath}
This is another reason why the invariance that we have discovered in this section can be considered as a generalization of the standard scale symmetry to which it reduces in the particular case $n=-2$. 

In the next section we will show that this symmetry manifests itself not only in the KvN formulation but also in the standard approach to classical mechanics. 

\section{Landau mechanical similarity}

A symmetry which in classical mechanics holds for every monomial potential, like the one of the previous section, was found long ago and presented by Landau in his book \cite{Landau}. In this section we want to prove that the transformations (\ref{transformations}) are just the KvN version of the transformations found by Landau. He realized that every monomial potential $\displaystyle V(r)=g \frac{r^n}{n}$ satisfies the equation $V(\alpha \vec{r}\,)=\alpha^n V(\vec{r}\,)$, so if we send
\begin{equation}
\left\{
\begin{array}{l}
\vec{r} \to \alpha \vec{r} \medskip \\
t \to \alpha^{1-n/2} t  \label{brr}
\end{array}
\right.
\end{equation}
the standard Lagrangian changes by an overall factor:
\begin{equation}
L=\frac{1}{2}\dot{r}^2-g\frac{r^n}{n}\, \longrightarrow \, \alpha^n\, L. \label{9bis}
\end{equation}
This implies that the classical equations of motion do not change under the transformations (\ref{brr}) which, consequently, can be considered a symmetry for the classical system. 
As we mentioned in the Introduction we will indicate this symmetry as LMS for Landau mechanical similarity. 
Under the transformations (\ref{brr}) the momenta $\displaystyle \vec{p}=\frac{\textrm{d}\vec{r}}{\textrm{d}t}$ change as follows: 
$\displaystyle \vec{p} \, \longrightarrow \, \alpha^{n/2} \vec{p}.$
If we write $\alpha=e^{\beta}$ and consider an infinitesimal $\beta$, then the variations of $t$ and of the phase space variables turn out to be:
\begin{equation}
\displaystyle \delta \vec{r}=\beta\, \vec{r}, \qquad \delta \vec{p} =\beta \,\frac{n}{2}\, \vec{p}, \qquad  \delta t=\beta \frac{2-n}{2}t. \label{tre}
\end{equation}
It is easy to realize from the manner $\vec{r}$ and $\vec{p}$ transform that, except for the inverse square potential ($n = -2$), the standard Poisson brackets $\{r_i,p_j\}=\delta_{ij}$  are not preserved by the transformations (\ref{tre}). This means that in the standard phase space formulation of classical mechanics the LMS cannot be implemented as a canonical transformation. 

We want now to prove that the transformations (\ref{transformations}) that we have found in the KvN space reproduce exactly the LMS tranformations of Eq. (\ref{tre}). Let us introduce a parameter $\tilde{\alpha}$ defined as: $\displaystyle \tilde{\alpha}\equiv
\frac{\beta(n-2)}{2}$, then Eq. (\ref{tre}) becomes:
\begin{equation}
 \delta \vec{r} = -\frac{2\tilde{\alpha}}{2-n} \, \vec{r}, \quad  \delta \vec{p}=-\frac{n\tilde{\alpha}}{2-n}\, \vec{p}, \quad \delta t=-\tilde{\alpha} t. \label{dieci} 
\end{equation}
Note that the transformations on $\vec{r}$ and $t$ above are exactly the same as the ones in (\ref{transformations}).
In the enlarged KvN space the momenta canonically conjugated to $\vec{r}$ and $\vec{p}$ are $\vec{\lambda}_r$ and $\vec{\lambda}_p$ respectively as one can notice from Eq. (\ref{conjugate}).
This gives us the possibility of implementing {\it canonically} in the enlarged space the transformations (\ref{dieci}), provided we transform the conjugate momenta $\vec{\lambda}$ with opposite signs w.r.t. the ones which appear in Eq. (\ref{dieci}), i.e.:
\begin{equation}
\delta \vec{\lambda}_r=\frac{2 \tilde{\alpha}}{2-n}\vec{\lambda}_r, \qquad  \delta \vec{\lambda}_p=\frac{n\tilde{\alpha}}{2-n}\vec{\lambda}_p. \label{tredici}
\end{equation}
By ``{\it canonically in the enlarged space}" we mean that the transformations of Eqs. (\ref{dieci}) and (\ref{tredici}) preserve the KvN commutators (\ref{conjugate}) or the associated extended Poisson brackets (epb)
\begin{equation}
\{r_i,\lambda_j\}_{\textrm{epb}}=\delta_{ij}, \qquad \{p_i,\lambda_{p_j}\}_{\textrm{epb}}=\delta_{ij}, \label{epb}
\end{equation}
which were introduced in Ref. \cite{gozzi}. Note that the request of having a canonical transformation in the enlarged space has generated in (\ref{tredici}) a transformation for $\vec{\lambda}_p$ identical to the one present in (\ref{transformations}). This proves that the transformations we found in (\ref{transformations}) are the KvN version of the LMS. This proves also that, while the LMS in $(\vec{r},\vec{p})$ space cannot be implemented canonically as shown in (\ref{tre}), this obstruction is removed in the enlarged KvN space. 

One last topic we want to present in this section is an extension of the analogy between scale symmetry and LMS.
It is known that the scale invariant inverse square potential is invariant also under special conformal transformations \cite{pino} and under an entire set of Virasoro charges \cite{carta}:
\begin{equation}
\displaystyle L_m=H\left(t+\frac{D_{0}}{H}\right)^{1+m}, \label{ccm}
\end{equation}
where $\displaystyle D_{0}=-\frac{pq}{2}$ and $\displaystyle H=\frac{p^2}{2}+\frac{g}{2q^2}$. 
These are the Noether charges associated with the infinitesimal time transformation \footnote{As particular cases, for $m=-1$ we get an infinitesimal time translation and $L_{\scriptscriptstyle -1}=H$; for $m=0$ we get a scale transformation and the Virasoro charge reproduces the usual dilation charge: $L_{0}=Ht-pq/2$.}   $t\rightarrow t-\epsilon t^{m+1}$. The $L_m$ of Eq. (\ref{ccm}) are conserved as a consequence of the following Poisson brackets $\{H,D_{0} \}_{\textrm{pb}}=H$. 

A natural question to ask is whether it is possible to find, also for the LMS invariant potentials analyzed in this paper,  further symmetries analog to the special conformal and the Virasoro algebras. The answer is yes. Using a notation analog to the one of Eq. (\ref{ccm}), let us call ${\cal D}_{0}$ the expression of the LMS charge of Eq.  (\ref{mscharge}) at time $t=0$. Combining ${\cal H}$ and ${\cal D}_{0}$ we can build an entire set of Virasoro charges given by:
\begin{equation}
\displaystyle {\cal L}_m={\cal H}\left( t+\frac{{\cal D}_0}{{\cal H}}\right)^{1+m}.  \label{vira}
\end{equation}
Using the extended Poisson brackets (\ref{epb}) we have that $\{{\cal H},{\cal D}_0\}_{\textrm{epb}}={\cal H}$ which implies that all the charges ${\cal L}_m$ of Eq. (\ref{vira}) are conserved under the evolution generated by ${\cal H}$, i.e. $\displaystyle \frac{\textrm{d}}{\textrm{d}t} {\cal L}_m=0$. 
The action of the classical path integral (\ref{gentilde}) turns out to be invariant under the transformations generated by ${\cal L}_m$ via the extended Poisson brackets (\ref{epb}), provided we transform time as follows: $\delta t =-\epsilon t^{m+1}$. Also in this case for $m=-1$ we get the invariance under infinitesimal time translations and the conserved charge (\ref{vira}) reduces to the Liouvillian ${\cal H}$. When $m=0$ we get instead the invariance of the action of the classical path integral (\ref{gentilde}) under the LMS transformations and the Virasoro charge (\ref{vira}) reduces to the LMS charge of Eq. (\ref{mscharge}). 

So we can conclude that also the LMS invariant potentials present an infinite set of other symmetries like the scale invariant potentials do \cite{carta}. 
A natural question to ask is whether these extra symmetries manifest themselves also in the standard formulation of classical mechanics, i.e., in the usual phase space $(\vec{r},\vec{p})\equiv \varphi$, or only in the extended phase space $(\varphi,\lambda)$ of the KvN formulation. To answer this question let us note that, among the ${\cal L}_m$, only ${\cal L}_0=t{\cal H}+{\cal D}_0$ and ${\cal L}_{-1}={\cal H}$ are linear in the variables $\lambda$. This implies that, once we apply them on the space $\varphi$ via the epb (\ref{epb}), we end up again in the space $\varphi$ 
\begin{displaymath}
\varphi \, \longrightarrow \, \varphi.
\end{displaymath}
This means that we can implement and see these symmetries even in the standard phase space (maybe in a non-canonical way, like the LMS). Acting instead with generators not linear in $\lambda$, like all the ${\cal L}_m$ (with $m\neq 0,-1$), the transformations on the space $\varphi$ will bring us into the $(\varphi,\lambda)$-space, as it is clear from Eq. (\ref{epb}), so
\begin{displaymath}
\varphi \, \longrightarrow \, (\varphi,\lambda).
\end{displaymath}
This means that these symmetries cannot be implemented and seen in the usual phase space $(\varphi)$ but only in the full KvN space $(\varphi,\lambda)$.

\section{Quantum mechanics in the KvN Hilbert space}

What we would like to understand in the next two sections is whether the LMS is preserved after quantization, i.e. whether the LMS can be considered as a symmetry also at the quantum level. For simplicity, we will limit ourselves to the one-dimensional case in which we have only one variable $q$, one variable $p$ and their associated momenta $\lambda_q$ and $\lambda_p$. The results can be easily generalized to higher dimensions.
As we have already seen in the previous sections, the LMS can be implemented as a {\it canonical} transformation only in the KvN space. 
So it seems natural to look for a corresponding quantum {\it unitary} transformation by implementing also quantum mechanics (QM) in the KvN Hilbert space. This is not
the Moyal formulation of QM \cite{moyal}, but something different explored in Ref. \cite{map}. In that paper one of us (D.M.) proved that, by defining on the KvN Hilbert space the following operators:
\begin{equation}
\hat{Q}\equiv \hat{q}-\frac{1}{2}\hbar\hat{\lambda}_p, \qquad \hat{P}\equiv \hat{p}+\frac{1}{2}\hbar\hat{\lambda}_q, \label{bopp1}
\end{equation}
one can reproduce the Heisenberg commutator $\left[ \hat{Q},\hat{P} \right]=i\hbar$ and the whole algebra of quantum observables by considering all the operators of the form $\displaystyle f(\hat{Q}, \hat{P})$ which are Hermitian under the KvN scalar product (\ref{scprod}). In particular, the quantum energy in the KvN space becomes the operator $H(\hat{Q},\hat{P})$ obtained by replacing the classical phase space variables $q$, $p$ with the operators $\hat{Q}$, $\hat{P}$ of Eq. (\ref{bopp1}). This $H(\hat{Q}, \hat{P})$ in general does not commute with the Liouvillian. Nevertheless, the quantum energy is conserved if we modify the Liouville equation as follows:
\begin{equation}
\displaystyle i\frac{\partial}{\partial t}|\psi\rangle=\hat{\cal G}|\psi \rangle, \quad 
\displaystyle \hat{\cal G}\equiv\frac{1}{\hbar}\left[H(\hat{Q},\hat{P})-H(\hat{\bar{Q}},\hat{\bar{P}})\right], \label{difference}
\end{equation}
where $\hat{\bar{Q}}$ and $\hat{\bar{P}}$ are the following operators:
\begin{equation}
\displaystyle \hat{\bar{Q}}\equiv \hat{q}+\frac{1}{2}\hbar\hat{\lambda}_p, \qquad \hat{\bar{P}}\equiv \hat{p}-\frac{1}{2}\hbar\hat{\lambda}_q. \label{bar}
\end{equation}
It is easy to realize that Eq. (\ref{difference}) goes into the Liouville equation when $\hbar \to 0$. 

The abstract KvN states $|\psi\rangle$ appearing in (\ref{difference}) can be represented on a basis of our choice. The one we will use from now on is made by the simultaneous eigenstates of the commuting operators $\hat{Q}$, $\hat{\bar{Q}}$ which we will indicate with $|Q,\bar{Q}\rangle$. The abstract states $|\psi \rangle$ then become $\psi(Q,\bar{Q})=\langle Q,\bar{Q}|\psi\rangle$.
The action of the generic quantum observable $\hat{F}\equiv(\hat{Q},\hat{P})$ on $\psi(Q,\bar{Q})$ is given by:
\begin{equation}
\displaystyle \hat{F}\psi(Q,\bar{Q})=f\left(Q,-i\hbar \frac{\partial}{\partial Q}\right) \psi(Q,\bar{Q}). \label{quantobs}
\end{equation}
If we consider the KvN Hilbert space as the tensor product of the Hilbert spaces spanned by the two basis $\{|Q\rangle\}$ and $\{|\bar{Q}\rangle\}$ respectively, then we can write the quantum observables as $\hat{F}\otimes \mathbb{I}$. 
This immediately tells us that, since we are describing quantum mechanics in a Hilbert space which is ``bigger" than the standard Hilbert space of quantum mechanics, there is a redundancy in the physical description. 
A way to remove this redundancy is to find a subspace of the whole KvN Hilbert space where the position $\hat{Q}$ and the momentum $\hat{P}$ {\it act irreducibly} (for details see Ref. \cite{map}).
This non-trivial subspace ${\bf H}_{\chi}$ can be built by making the product of any normalizable wave function $\psi$ in $Q$ with a fixed wave function in $\bar{Q}$, which we indicate with $\chi(\bar{Q})$: \footnote{This is quite similar to the procedure introduced by M. Ban in Ref. \cite{ban}.}
\begin{equation}
\displaystyle {\bf H}_{\chi}= \left\{ \psi(Q)\chi(\bar{Q}) \, \textrm{with} \, \int \textrm{d}Q 
\textrm{d}\bar{Q} \, |\psi(Q)|^2 |\chi(\bar{Q})|^2 =1\right\}. \label{subspace}
\end{equation}
Because $\chi$ is fixed, ${\bf H}_{\chi}$ is isomorphic to the standard Hilbert space of quantum mechanics with the standard scalar product:
\begin{displaymath}
\displaystyle \langle \psi|\psi^{\prime}\rangle =\int \textrm{d}Q\, \psi^*(Q)\psi^{\prime}(Q),
\end{displaymath}
which is naturally induced by the scalar product (\ref{scprod}). Note that all the Hilbert subspaces $H_{\chi}$, $H_{\chi^{\prime}}$, $H_{\chi^{\prime\prime}},\cdots$, obtained by changing the fixed function $\chi$ are isomorphic to each other. The quantum observables act on the KvN states as given by Eq. (\ref{quantobs}), so it is easy to realize that they map vectors of (\ref{subspace}) into vectors of (\ref{subspace}). 
Finally, note that, when we restrict ourselves to the subspace ${\bf H}_{\chi}$ (or to any of the equivalent subspaces), we have from Eq. (\ref{difference}) that the function $\psi(Q)$ evolves with the usual Schr\"odinger equation:
\begin{displaymath}
\displaystyle i\frac{\partial}{\partial t}\psi(Q)=\frac{1}{\hbar}H(\hat{Q},\hat{P})\psi(Q).
\end{displaymath}
For more details on this KvN realization of QM we refer the reader to Ref. \cite{map}.

Now that we have formulated quantum mechanics in the KvN Hilbert space let us go back to the LMS symmetry. The natural question to ask in general is the following: how can we implement a symmmetry at the quantum level in this framework?  
As we have already seen, the operator which generates the quantum evolution is given by Eq. (\ref{difference}). If we use the definitions (\ref{bopp1}) and (\ref{bar}) then it is easy to see that the operator of Eq. (\ref{difference}) can be written as: 
\begin{eqnarray}
\displaystyle \hat{\cal G}&=&\sum_{j=0}^{\infty} \frac{\hbar^{2j}}{2^{2j}(2j+1)!} \lambda_{a_1}\cdots \lambda_{a_{2j+1}}\omega^{a_1b_1}\cdots \omega^{a_{2j+1}b_{2j+1}}\nonumber \\ 
&& \partial_{b_1}\cdots 
\partial_{b_{2j+1}}H(q,p).  \label{corracca}
\end{eqnarray}
This is basically the Liouville operator modified by an infinite set of corrections in increasing powers of $\hbar$. 
The change from the Liouville operator to $\hat{\cal G}$, which we performed for that particular canonical transformation which is the time evolution, must be done for any canonical transformation. What we mean is the following: if the function $C(q,p)$ generates via the Poisson brackets a certain transformation in the standard phase space formulation of classical mechanics, then the same transformation is implemented in the KvN space via the Hamiltonian vector field \cite{quantpro} associated with $C(q,p)$, i.e. via $\hat{\cal C}=\lambda_a\omega^{ab}\partial_bC(q,p)$ which plays the same role that the Liouvillian played for the time evolution \cite{gozzi}. The operator which generates the same transformation at the quantum level can be written in the same form of the operator $\hat{\cal G}$ of evolution of Eq. (\ref{difference}) but with the Hamiltonian $H$ replaced by the function $C$:
\begin{equation}
\displaystyle {\cal C}_{\hbar}\equiv\frac{1}{\hbar} \left[C(\hat{Q},\hat{P})-C(\hat{\bar{Q}},\hat{\bar{P}})\right]. \label{generalized0}
\end{equation}
This is equivalent to modify the Hamiltonian vector field with the corrections in $\hbar$ given by the following expression:
\begin{eqnarray}
 \displaystyle \hat{\cal C}_{\hbar}&=&\sum_{j=0}^{\infty} \frac{\hbar^{2j}}{2^{2j}(2j+1)!} \lambda_{a_1}\cdots \lambda_{a_{2j+1}}\omega^{a_1b_1}\cdots \omega^{a_{2j+1}b_{2j+1}}\nonumber\\
&&\partial_{b_1}\cdots \partial_{b_{2j+1}}C(q,p). \label{generalized}
\end{eqnarray}
When we send $\hbar\to 0$ we have that $\hat{\cal C}_{\hbar} \to \hat{\cal C}=\lambda_a\omega^{ab}\partial_bC$, i.e. we get just the Hamiltonian vector field associated with the charge $C$, which generates the symmetry at the classical level. 
The expression (\ref{generalized}) has appeared before in the literature \cite{moyal} but not in a Hilbert space context. 
Before concluding this section, let us notice that, since $\hat{Q}$ and $\hat{P}$ commute with $\hat{\bar{Q}}$ and $\hat{\bar{P}}$, the variation induced by $\hat{\cal C}_{\hbar}$ on a function of $\hat{Q}$ and $\hat{P}$ is again a function of $\hat{Q}$ and $\hat{P}$, see eq. (\ref{generalized0}), so the transformation does not bring us outside the space of the observables $f(\hat{Q},\hat{P})$. 

Unfortunately things become more subtle when we consider the LMS symmetry. In fact, as we have seen in Sec. III, the transformations of the LMS are not canonical in the standard phase space of classical mechanics, so there is no function $C(q,p)$ which generates the transformations via the usual Poisson brackets. Consequently, we have no $C(q,p)$ to put into the definition (\ref{generalized}) of the charge $\hat{\cal C}_{\hbar}$ which generates the transformations at the quantum level, so we have to use a different strategy.

\section{Mechanical similarity at the quantum level}

Let us start by considering the LMS symmetry for the harmonic oscillator. In this case the Hamiltonian $H(q,p)$ is quadratic in $q$ and $p$, so all the corrections in $\hbar$ in the operator $\hat{\cal G}$ of Eq. (\ref{corracca}) disappear. This means that the Liouvillian itself generates the evolution at the quantum level. Let us also note that, as the charge of mechanical similarity of Eq. (\ref{msosc}) commutes with the Liouvillian, we can say that it is a conserved charge both at the classical and the quantum level, so we think that it may be this same charge which generates the quantum LMS transformation. The associated unitary operator will be 
\begin{equation}
\displaystyle U=\exp \left[ i\alpha(\lambda_qq+p\lambda_p)\right]. \label{unit}
\end{equation}
The reader may not be convinced that this is the full quantum operator and that $\hbar$-corrections should be present. We shall show later on for the general case that $\hbar$-corrections will not modify our conclusions. 
The transformations induced by $U$ on the quantum position $\hat{Q}$ and the quantum momentum $\hat{P}$ are:
\begin{eqnarray*}
&& \displaystyle U\hat{Q}U^{-1}=\textrm{sinh} \, \alpha \,\hat{\bar{Q}}+\textrm{cosh}\, \alpha \, \hat{Q}, \nonumber \\
&& \displaystyle U\hat{P}U^{-1}=\textrm{sinh}\, \alpha \,\hat{\bar{P}}+\textrm{cosh} \,\alpha \,\hat{P}.
\end{eqnarray*}
From the previous equation we see that, by applying the transformations of the LMS on the operators $\hat{Q}$ and $\hat{P}$, we get linear combinations not only of $\hat{Q}$ and $\hat{P}$, but also of $\hat{\bar{Q}}$, $\hat{\bar{P}}$. In general, when we apply the LMS transformations (\ref{unit}) on a QM observable, which is a Hermitian operator $f(\hat{Q},\hat{P})$, we will get a new operator which depends also on $\hat{\bar{Q}}$ and $\hat{\bar{P}}$, differently than what happened in the case of transformations of the type (\ref{generalized}).
That means that the LMS transformations bring us outside the space of the quantum observables. The same happens for the physical states of the theory.
In fact, let us rewrite the unitary transformation (\ref{unit}) in terms of $Q$ and $\bar{Q}$:
\begin{equation}
\displaystyle U=\exp \left[ \alpha \left(\bar{Q}\frac{\partial}{\partial Q}+Q\frac{\partial}{\partial \bar{Q}}\right)\right]. \label{unit1}
\end{equation}
For an infinitesimal $\alpha$ we have that $U$ can be rewritten as the following abstract operator in the KvN Hilbert space:
\begin{equation}
\displaystyle U={\mathbb I}\otimes {\mathbb I}+\frac{i\alpha}{\hbar}\left[ \hat{P}\otimes \hat{\bar{Q}}- \hat{Q}\otimes \hat{\bar{P}}\right]. \label{unit2}
\end{equation}
Let us now apply the unitary transformation (\ref{unit2}) on the states belonging to the Hilbert space of quantum mechanics ${\bf H}_{\chi}$, i.e. on the states of the form $\psi(Q)\chi(\bar{Q})$, with $\chi(\bar{Q})$ fixed \cite{map}. 
From Eqs. (\ref{unit1}) and (\ref{unit2}) we see that $U$ contains explicitly operators which act on the Hilbert space spanned by $\{ | \bar{Q} \rangle \}$, so
when we apply the transformation $U$ on a state $\psi(Q)\chi(\bar{Q})$ we obtain that the form of the state $\chi(\bar{Q})$ gets changed. Not only, but in general we get a wave function which is not separable anymore, so we get a state which does not belong to any of the equivalent subspaces of KvN space which are isomorphic to the Hilbert space ${\bf H}_{\chi}$ of quantum mechanics. 

These considerations can be easily generalized to an arbitrary monomial potential. In this case at the classical level the LMS in the KvN space is generated by the following unitary operator derived from (\ref{mscharge}):
\begin{eqnarray}
\displaystyle U&=&\exp \Bigl[ i\alpha\Bigl( t\hat{\cal H}-\frac{1}{2-n}(\lambda_qq+q\lambda_q)
\nonumber \\
&& \displaystyle -\frac{n}{2(2-n)}(\lambda_pp+p\lambda_p)\Bigr)\Bigr],
\label{unitop}
\end{eqnarray}
which depends explicitly on the operator of evolution $\hat{{\cal H}}$. When we implement quantum mechanics in the KvN space we know that we have to replace the Liouvillian $\hat{\cal H}$ with the operator $\hat{\cal G}$ of Eq. (\ref{corracca}). Since the classical Liouvillian appears in the classical charge of mechanical similarity, the same replacement mentioned above has to be performed within the unitary operator (\ref{unitop}) which implements the LMS. 
Furthermore let us keep open the possibility of modifying the part of the operator $U$ which does not depend on time $t$ with corrections in $\hbar$. Consequently, the operator which should generate mechanical similarity at the quantum level is, modulo further corrections in $\hbar$, the following one:
\begin{eqnarray}
\displaystyle U&=& \exp \Bigl[ i\alpha\Bigl( t\hat{\cal G}-\frac{1}{2-n}({\lambda}_q{q}+{q}{\lambda}_q)\nonumber \\
\displaystyle && -\frac{n}{2(2-n)}({\lambda}_p{p}+{p}{\lambda}_p)+O(\hbar)\Bigr)\Bigr]. \label{(A)}
\end{eqnarray}
The infinitesimal transformations induced by $U$ on the quantum position and momentum are, modulo terms of order $\hbar$,: 
\begin{equation}
\begin{array}{l}
\displaystyle U \hat{Q} U^{-1}=\alpha t \hat{P} -\frac{2}{2-n}\alpha \hat{Q}-\frac{n+2}{2(2-n)}\alpha (\hat{\bar{Q}}-\hat{Q})
 \medskip \\
\displaystyle U \hat{P} U^{-1}=- \alpha t g\hat{Q}^{n-1}-\frac{n}{2-n}\alpha \hat{P}-\frac{n+2}{2(2-n)}\alpha(\hat{\bar{P}}-\hat{P})
\label{outside}
\end{array} 
\end{equation}
The previous equations tell us that, except in the case of an inverse square potential ($n=-2$), the LMS transformations turn $\hat{Q}$ and $\hat{P}$ into combinations of not only $\hat{Q}$ and $\hat{P}$ but also of $\hat{\bar{Q}}$ and $\hat{\bar{P}}$. This implies that when we apply the transformations to the physical observables, i.e. $f(\hat{Q},\hat{P})$, we end up with functions that are not observables anymore  because they depend also on $\hat{\bar{Q}}$ and $\hat{\bar{P}}$. Let us notice that this happens even if we add corrections in $\hbar$ as we did in the operator $U$ of Eq. (\ref{(A)}). These corrections in fact cannot cancel the $(\hat{\bar{Q}},\hat{\bar{P}})$ terms in (\ref{outside}) which are already present at $\hbar=0$. Since the LMS brings every quantum observable outside the space of quantum observables and the same happens for the physical states, we conclude that the LMS cannot be implemented at the quantum level at least within the KvN space.

Of course, similar problems in implementing the LMS at the quantum level are present also in more standard formulations of quantum mechanics. For example, let us try to realize the LMS via a unitary transformation $\displaystyle U=\exp\left[ i\tilde{\alpha}\hat{A}/\hbar\right]$ acting on the standard Hilbert space of quantum mechanics. For an infinitesimal $\tilde{\alpha}$ we get: $\displaystyle {U}=\mathbb{I}+i \tilde{\alpha}\hat{A}/\hbar$. 
So let us ask ourselves whether, for a particular choice of the operator $U$ or, equivalently, of the operator $\hat{A}$, the transformations
\begin{displaymath}
\displaystyle \hat{q}^{\prime}=U\hat{q}U^{-1}, \qquad \hat{p}^{\prime}=U\hat{p}U^{-1}
\end{displaymath}
reproduce exactly the LMS transformations on the operators $\hat{q}$ and $\hat{p}$, which, from Eq. (\ref{dieci}), are: 
\begin{equation}
\left\{
\begin{array}{l}
\displaystyle \delta \hat{q}= -\frac{2}{2-n}\tilde{\alpha} \hat{q}+\tilde{\alpha} t \frac{\textrm{d}}{\textrm{d}t} \hat{q}, \medskip \\
\displaystyle \delta \hat{p}= -\frac{n}{2-n}\tilde{\alpha} \hat{p}+\tilde{\alpha} t \frac{\textrm{d}}{\textrm{d}t}\hat{p}. \label{standard}
\end{array}
\right.
\end{equation}
If we neglect terms in $\tilde{\alpha}^2$ we have:
\begin{displaymath}
\displaystyle U\hat{q}U^{-1}=\hat{q}+\frac{i\tilde{\alpha}}{\hbar}\left[\hat{A},\hat{q}\right],
\qquad U\hat{p}U^{-1}=\hat{p}+\frac{i\tilde{\alpha}}{\hbar}\left[\hat{A},\hat{p}\right].
\end{displaymath}
To reproduce the terms of Eq. (\ref{standard}) which depend explicitly on time $t$ we are forced to consider an operator $\hat{A}$ of the form $\hat{A}=t\hat{H}+\hat{A}_0$. The operator $\hat{A}_0$ is determined once we succeed in satisfying the following commutators with $\hat{q}$ and $\hat{p}$:
\begin{equation}
\displaystyle \frac{i}{\hbar} \left[\hat{A}_0,\hat{q}\right]=-\frac{2}{2-n}\hat{q},
\qquad \frac{i}{\hbar} \left[\hat{A}_0, \hat{p}\right]=-\frac{n}{2-n}\hat{p}. \label{90}
\end{equation}
This implies that $\hat{A}_0$ must have the form $\hat{A}_0=\alpha \hat{q}\hat{p}$.
In particular, the first equation tells us that $\displaystyle \tilde{\alpha}=-\frac{2}{2-n}$ and the second that $\displaystyle \tilde{\alpha}=\frac{n}{2-n}$. This means that, unless we consider the case $n=-2$ (in which the LMS reduces to a scale transformation), there does not exist any operator $\hat{A}_0$ which satisfies Eq. (\ref{90}). In other words, it is impossible to implement the LMS via a unitary operator acting on the standard Hilbert space formulation of quantum mechanics.

The problems in realizing the LMS at the quantum level can be understood in a more intuitive way if we adopt the old Bohr-Sommerfeld quantization rules. In fact, remember that at the classical level the LMS can be considered a symmetry because it just rescales the Lagrangian (\ref{9bis}), so it leaves unchanged the form of the equations of motion. This implies that the LMS maps a solution of the classical equations of motion into another solution of the same equations. For example in the case of a harmonic oscillator it maps an ellipse in phase space into another ellipse in phase space and if the transformation is infinitesimal it will map an ellipse into another one infinitesimally ``close" to it. Of course, things change when we consider quantum mechanics. In this case in fact the Bohr-Sommerfeld quantization rules impose that only some of the ``trajectories" are allowed: 
\begin{equation}
\displaystyle \oint \textrm{d}x\, p=\left(\bar{n}+1/2\right)h, \qquad \quad \bar{n}\in \mathbb{N}. \label{bohr}
\end{equation}
So when we apply an infinitesimal LMS transformation given by Eq. (\ref{tre}), we have that the LHS of (\ref{bohr}) changes by an arbitrary small quantity (except for $n=-2$). 
This infinitesimal change, as $\bar{n}$ is an integer, cannot be matched by a {\it discrete} change of $\bar{n}$ on the RHS of (\ref{bohr}). The only way out would be the possibility to change infinitesimally $h$ but we know that QM does not allow that. The reader may wonder that this reasoning of ours could be applied to any infinitesimal symmetry and not just to the LMS. This is not true. In fact, it is only the LMS, with its non-canonical form (\ref{tre}), that changes the LHS of (\ref{bohr}). 

Another way to realize that the LMS cannot be implemented at the QM level is to turn to the standard path integral \cite{fey} formulation of QM which we will briefly indicate with its generating functional:
\begin{equation}
Z=\int {\mathcal D}q{\mathcal D}p \, \exp\left[ \frac{i}{\hbar}S\right]. \label{29bis}
\end{equation}
From the manner the $L$ of Eq. (\ref{9bis}) changes under the LMS (\ref{brr}) we get that the action S in (\ref{29bis}) changes as
\begin{displaymath}
\displaystyle S\, \longrightarrow\, \alpha^{(1+n/2)} S.
\end{displaymath}
This rescale can be compensated in the $Z$ of (\ref{29bis}) only by a change in $\hbar$. In fact, even a change in the measure $\int {\cal D}q{\cal D}p$ cannot compensate the rescale of $S$. The reason is because the change induced by (\ref{tre}) in the measure does not depend on the potential (except for the dependence on $n$) while the rescale of $S$ pulls in the entire form of the potential with its dependence not only on $n$ but also on the coupling constant $g$ appearing in (\ref{9bis}). So we conclude that only a rescale of $\hbar$ would make the LMS a symmetry at the QM level. We feel that the LMS, with its connection to a rescaling of $\hbar$, is quite unique and it may play a role in the interface between classical mechanics (CM) and QM. One drawback of the LMS is that it is a symmetry of only the monomial potentials, so it cannot play a {\it universal} role in the interplay between CM and QM. The research we are now pursuing is to find a generalization of the LMS valid for any interaction, that means a transformation which rescales the action for any potential. This would be a {\it universal} symmetry which is {\it never} implementable in QM (because of $\hbar$) but {\it always} present in CM and so it would really mark the border between CM and QM. Some work has already been done in this direction \cite{deotto}. The price that one seems to pay in order to get a {\it universal} symmetry is that the transformation does not act on time $t$, like (\ref{brr}), but on some Grassmannian partners of time \cite{gozzi}-\cite{deotto} whose physical meaning is not yet clear. We are now trying to figure out how that symmetry \cite{deotto} could emerge in the standard formulation of CM and QM. 

\section{Connection with Newton-equivalent Hamiltonians}

The reader familiar with this kind of topics may like to compare what we did in this paper with what people have done in the sector of ``Newton-equivalent systems" and their quantization \cite{cal}. First of all, let us briefly review Ref. \cite{cal}. If we consider a particle of mass $m$ in a potential $V$ then the Newtonian equations of motion are given by $\displaystyle m\ddot{q}=-V^{\prime}(q)$. These equations of motion can be derived from the {\it st}andard Lagrangian:
\begin{equation}
\displaystyle L_{\textrm{st}}=m\frac{\dot{q}^2}{2}-V(q)   \label{lagst}
\end{equation}
or from any of the following equivalent Lagrangians:
\begin{equation}
\displaystyle L_{\gamma}=\gamma\left( m\frac{\dot{q}^2}{2}-V(q)\right).   \label{laggen}
\end{equation}
Note that we get different momenta canonically conjugated to $q$, according to the different values of $\gamma$:
\begin{equation}
\displaystyle p_{\gamma}=\frac{\partial L_{\gamma}}{\partial \dot{q}}=\gamma m\dot{q}=\gamma p. \label{palfa}
\end{equation}
Performing the Legendre transform we get a whole set of equivalent Hamiltonians labeled by $\gamma$:
\begin{equation}
\displaystyle H_{\gamma}(p_{\gamma},q;\gamma)=\frac{p_{\gamma}^2}{2\gamma m}+\gamma V(q).
\label{hamgen}
\end{equation}
In particular, for $\gamma=1$ Eq. (\ref{hamgen}) reduces to:
\begin{equation}
\displaystyle H_{\textrm{st}}(p,q)=\frac{p^2}{2m}+V(q),  \label{hamstand}
\end{equation}
which is the Hamiltonian associated with the standard Lagrangian of Eq. (\ref{lagst}). 
In Ref. \cite{cal} the following Poisson brackets is imposed between $q$ and $p_{\gamma}$: 
\begin{equation}
\{q,p_{\gamma}\}=1.  \label{poissbr}
\end{equation}
Applying then the standard quantization rules on (\ref{poissbr}) the momentum becomes an operator independent of $\gamma$, i.e. $\displaystyle \hat{p}_{\gamma}=-i\hbar\frac{\partial}{\partial q}$, while the Hamiltonian becomes the following operator: 
\begin{displaymath}
\displaystyle \hat{H}_{\gamma}=-\frac{\hbar^2}{2\gamma m}\frac{\partial^2}{\partial q^2}+\gamma V(q)
\end{displaymath}
which depends explicitly on the value of $\gamma$. This implies that both the eigenvalues and the eigenfunctions of $\hat{H}_{\gamma}$ depend explicitly on $\gamma$. 
So at the {\it quantum level} the dynamics given by $\hat{H}_{\gamma}$ is different than the one given by $\hat{H}$ while it was the same at the {\it classical level}. 
We can summarize what Calogero et al. did in Ref. \cite{cal} in the following picture:

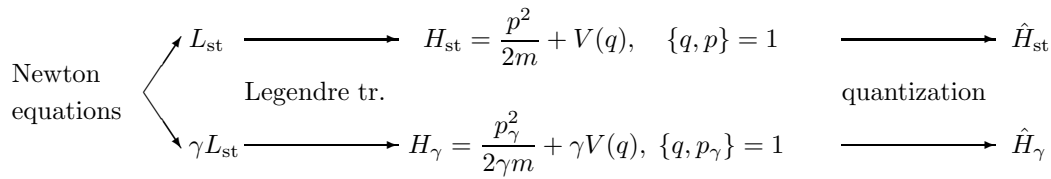
\begin{widetext}
\hspace{1.5cm} \begin{picture}(200,100)
\put(10,57){Newton}
\put(10,43){equations}
\put(60,53){\vector(2,3){14}} \put(77,70){$\displaystyle L_{\textrm{st}}$}
\put(98,50){Legendre tr.}
\put(160,70){$\;\; \displaystyle H_{\textrm{st}}=\frac{p^2}{2m}+V(q), \quad \{q,p\}=1$}
\put(160,30){$\displaystyle H_{\gamma}=\frac{p_{\gamma}^2}{2\gamma m}+\gamma V(q),
\; \{q,p_{\gamma}\}=1$}
\put(98,73){\vector(1,0){58}}  \put(324,73){\vector(1,0){58}}
\put(60,53){\vector(2,-3){14}}  \put(77,30){$\displaystyle \gamma L_{\textrm{st}}$}
\put(324,33){\vector(1,0){58}} \put(388,30){$\hat{H}_{\gamma}$}
\put(388,70){$\hat{H}_{\textrm{st}}$}
\put(98,33){\vector(1,0){58}}
\put(324,50){quantization}
\end{picture}
\end{widetext}

In our formalism, a part from restricting $V(q)$ to be a monomial, we have 
in common with Ref. \cite{cal} the fact that our Lagrangian also rescales by a factor (\ref{9bis}). In our case the transformation from one Lagrangian to the other was obtained via some explicit transformation on $q$, $p$ and $t$ (the LMS) while this was not the case in \cite{cal}. There they just postulated the two different Lagrangians or Hamiltonians, without connecting them via a transformation. As we have the explicit transformation we need to implement it also at the canonical level and not just impose some Poisson brackets between the $p$ and the $q$. What we get after the LMS transformation is a Hamiltonian and a canonical structure different than the one of Ref. \cite{cal}. This is outlined in the scheme below: 

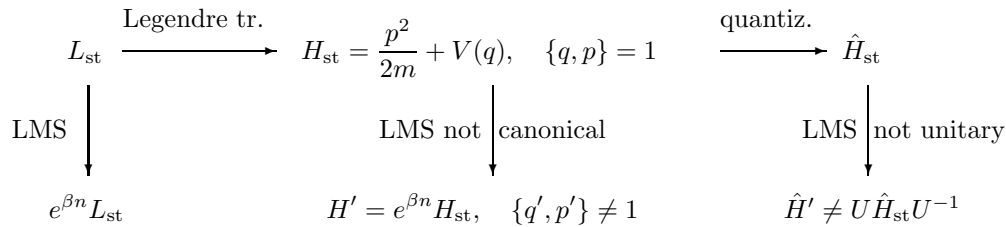
\begin{widetext}
\begin{picture}(200,100)
\put(56,40){LMS}
\put(85,60){\vector(0,-1){34}} \put(77,70){$\displaystyle L_{\textrm{st}}$}
\put(98,82){Legendre tr.}
\put(324,82){quantiz.}
\put(160,70){$\;\; \displaystyle H_{\textrm{st}}=\frac{p^2}{2m}+V(q), \quad \{q,p\}=1$}
\put(175,10){$\displaystyle H^{\prime}=e^{\beta n} H_{\textrm{st}}, \quad \{q^{\prime},p^{\prime}\}\neq 1$}
\put(98,73){\vector(1,0){58}}  \put(324,73){\vector(1,0){40}}
 \put(70,10){$\displaystyle e^{\beta n} L_{\textrm{st}}$}
 \put(348,10){$\hat{H}^{\prime}\neq U\hat{H}_{\textrm{st}}U^{-1}$}
 \put(355,40){LMS} \put(382,40){not unitary}
\put(370,70){$\hat{H}_{\textrm{st}}$}
\put(238,60){\vector(0,-1){34}} \put(195,40){LMS not}  \put(240,40){canonical}
\put(380,60){\vector(0,-1){34}}
\end{picture}
\medskip 
\end{widetext}

In the first column above the LMS transformations (\ref{tre}) on the position $q$ and the momentum $p$ map the Lagrangian (\ref{lagst}) into the Lagrangian (\ref{laggen}) with $\gamma=e^{\beta n}$. After having performed the Legendre transform on $L_{\textrm{st}}$ the same LMS would map 
$\displaystyle H_{\textrm{st}}=\frac{p^2}{2m}+V(q)$ into\break $H^{\prime}=e^{\beta n} H_{\textrm{st}}$, where $n$ is the exponent of the monomial potential. If we consider an infinitesimal parameter $\beta$ and we disregard terms of order $\beta^2$ then the fundamental Poisson brackets $\{ q,p\}=1$ are mapped into 
\begin{displaymath}
\displaystyle \{ q^{\prime},p^{\prime} \}=\left\{q+\beta q, p+\beta\frac{n}{2}p\right\}=
1+\beta\left(1+\frac{n}{2}\right) 
\end{displaymath}
which are different than the ones imposed between $q$ and $p_{\gamma}$ of Ref. \cite{cal}. 
Transformations which change the Poisson structure, like ours do, are known in the literature: they are called {\it canonical} but {\it not completely canonical} in Ref. \cite{wintner}, or {\it conformal symplectic} transformations in Ref. \cite{basart}. If we now try to quantize in the standard way we get that the LMS cannot be implemented via a unitary transformation (see previous section and the last arrow in the scheme above). So, summarizing, by applying our LMS transformation we connect the two Lagrangians as in Ref. \cite{cal}, nevertheless if we start from $H_{\textrm{st}}$ and $\{q,p\}=1$ and we apply a LMS, we do not get the Hamiltonian $H_{\gamma}$ and the Poisson brackets $\{q,p_{\gamma}\}=1$ as in Ref. \cite{cal}. We get instead the Hamiltonian $H^{\prime}$ and the Poisson brackets $\displaystyle \{q^{\prime},p^{\prime}\}=1+\beta \left(1+\frac{n}{2}\right)\neq 1$. So the ``canonical" structure obtained in the procedure \cite{cal} and ours are totally different. As a consequence also the quantum structure is different. While the authors of \cite{cal}, having a canonical structure in $H_{\gamma}$, can proceed to quantize, we have first to pass to a formalism in which the LMS can be implemented canonically. That is the KvN formalism. This is summarized in the first row of the scheme below:

\begin{widetext}
\hspace{-2cm}
\begin{picture}(200,100)
\put(324,82){quantiz.}
\put(160,70){$\;\; \displaystyle {\cal H}_{\textrm{st}}=\lambda_a\omega^{ab}\partial_bH_{\textrm{st}}, \quad \{\varphi,\lambda\}=1$}
\put(175,10){$\displaystyle {\cal H}^{\prime}=e^{\tilde{\alpha}} {\cal H}_{\textrm{st}}, \quad \{\varphi,\lambda\}=1 $}
\put(324,73){\vector(1,0){40}}
 \put(358,10){${H}_{\textrm{st}}(\hat{Q}^{\prime},\hat{P}^{\prime})$}
 \put(352,-3){not observable}
 \put(355,35){LMS} \put(382,35){unitary}
\put(370,70){${H}_{\textrm{st}}(\hat{Q},\hat{P})$}
\put(230,55){\vector(0,-1){34}} \put(205,35){LMS \, canonical}
\put(380,55){\vector(0,-1){34}}
\end{picture}

\bigskip 
\end{widetext}

If we now quantize starting from the KvN formalism we get that the LMS brings us outside the physical Hilbert space and the space of observables (see Sec. V and the last arrow of the scheme above). So the symmetry cannot be implemented at the QM level. To summarize, the picture we get at the QM level is different from the one of Ref. \cite{cal} just because the ``canonical" structure is different. We are forced on this canonical and quantum structure by the fact that we have an explicit form of the transformation which rescales the Lagrangian while that is not the case in Ref. \cite{cal}. 

We can conclude that while in Ref. \cite{cal} the ``symmetry" of rescaling the Lagrangian cannot be {\it maintained} at the quantum level, in our case it cannot even be {\it implemented}.

\section*{Acknowledgments}
We would like to thank G. Marmo and M. Reuter  for useful discussions. This research has been supported by grants from INFN, MIUR and the University of Trieste.

\end{document}